\documentclass[]{spie}

\usepackage{amsmath,amsfonts,amssymb}
\usepackage{graphicx}
\usepackage[colorlinks=true, allcolors=blue]{hyperref}
\usepackage{bm}

\title{Theory of spin pumping and inverse Rashba-Edelstein effect in a two-dimensional electron gas}

\author[a]{Masaki Yama}
\author[b,c,d,e]{Mamoru Matsuo}
\author[a]{Takeo Kato}
\affil[a]{Institute for Solid State Physics, University of Tokyo, 5-1-5 Kashiwanoha, Kashiwa, Japan}
\affil[b]{Kavli Institute for Theoretical Sciences, University of Chinese Academy of Sciences, Beijing, 100190, China}
\affil[c]{CAS Center for Excellence in Topological Quantum Computation, University of Chinese Academy of Sciences, Beijing 100190, China}
\affil[d]{Advanced Science Research Center, Japan Atomic Energy Agency, Tokai, 319-1195, Japan}
\affil[e]{RIKEN Center for Emergent Matter Science (CEMS), Wako, Saitama 351-0198, Japan}

\authorinfo{Further author information: (Send correspondence to Takeo Kato)\\Takeo Kato: E-mail: kato@issp.u-tokyo.ac.jp, Telephone: 81 4 7136 3255}

\begin{document} 
\maketitle

\begin{abstract}
We theoretically investigate spin transport in a junction system composed of a ferromagnetic insulator (FI) and a two-dimensional electron gas (2DEG) with both Rashba and Dresselhaus spin-orbit interactions. We briefly present our findings on spin pumping into the 2DEG under microwave irradiation and subsequent current generation via the inverse Rashba-Edelstein effect. We highlight the crucial role of vertex corrections in the theoretical description of these effects.
\end{abstract}


\section{INTRODUCTION}
\label{sec:intro} 

Spin pumping has long been used as a method of injecting spins into various materials~\cite{Tserkovnyak2002,Tserkovnyak2005,Mizukami2001,Saitoh2006,Kajiwara2010,Hellman2017}. The amount of spin injection is evaluated by the increase in Gilbert damping measured in ferromagnetic resonance (FMR) experiments~\cite{Han2020}. Spin pumping experiments can also be utilized for the measurement of spin excitation in target systems~\cite{Hirobe2019,Kato2019,Yamamoto2021,Ominato2022a,Ominato2022b,Funato2022,Fukuzawa2023,Haddad2023}. The signal of spin pumping is detected by a current or voltage induced in an adjacent metal through spin-charge conversion effects such as the spin Hall effect.

Recently, spin-charge conversion in two-dimensional electron systems via the inverse Rashba-Edelstein effect has gained attention in the development of spintronic devices~\cite{Shen2014a, Manchon2015, Soumyanarayanan2016}. By combining spin pumping with the inverse Rashba-Edelstein effect, charge current generation has been studied in materials such as the Ag/Bi interface and SrTiO${}_3$ heterostructures.\cite{Sanchez2013,Lesne2016, Song2017}

\begin{figure}[tb]
\centering
\includegraphics[width=120mm]{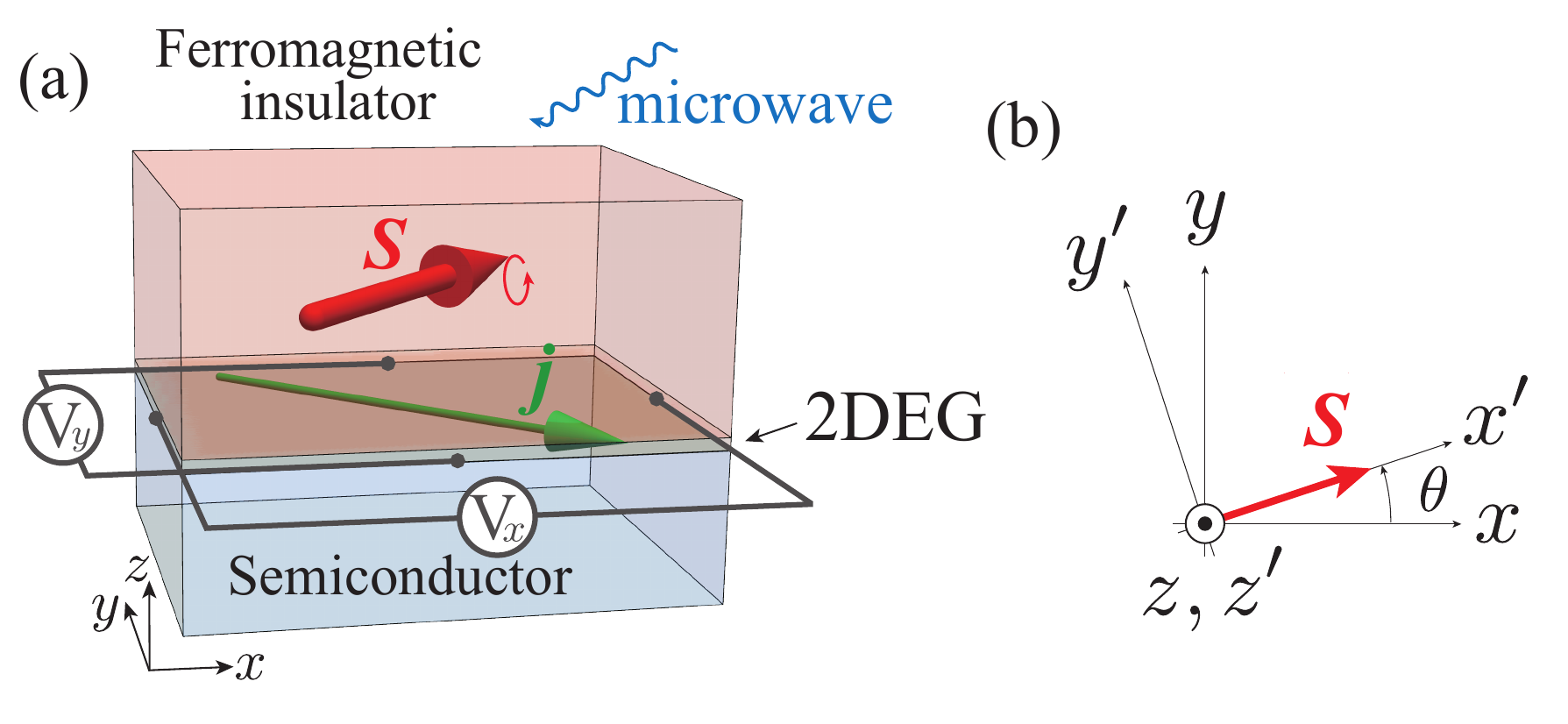}
\caption{(a) A junction system considered in our study. The red arrow, $\bm{S}$, indicates the spontaneous spin polarization of the ferromagnetic insulator (FI), which induces spin precession under microwave irradiation. The green arrow, $\bm{j}$, represents the current density generated by the inverse Rashba-Edelstein effect. (b) Coordinate transformation from a laboratory coordinates $(x, y, z)$ to a new coordinates $(x', y', z')$ fixed to the spin polarization of the FI, $\bm{S}$, in the absence of microwave irradiation, which is indicated by the red arrow.}
\label{fig:setup}
\end{figure}

In our study, we consider spin pumping in a junction composed of a ferromagnetic insulator (FI) and a two-dimensional electron gas (2DEG), as shown in Fig.~\ref{fig:setup}. As a prototype system with spin-orbit interaction, we consider conduction electrons with a parablic band dispersion and both of the Rashba and Dresselhaus spin-orbit interactions, which have been extensively studied in semiconductor heterostructures.\cite{Bernevig2006,Kohda2017,Ganichev2004,Trushin2007} 
We theoretically formulate the increase of Gilbert damping in spin pumping experiments using the diagrammatic technique. Furthermore, we develop a formulation for the inverse Rashba-Edelstein effect in 2DEG driven by spin pumping based on the Boltzmann equation. 
In both formulations, we emphasize the importance of considering the conservation law through vertex corrections or the full solution of the Boltzmann equation without using the relaxation-time approximation.
We should note that our formulation can be extended for application to various materials with spin-splitting bands in the presence of the spin-orbit interaction.

\section{MODEL}
\label{sec:model} 

We consider a 2DEG/FI junction (see Fig.~\ref{fig:setup}) based on a microscopic model. Let us first introduce a model Hamiltonian for 2DEG. The kinetic energy of the 2DEG is described by the following Hamiltonian:
\begin{align}
{\cal H}_{\rm kin}&=\sum_{\bm{k},\sigma,\sigma'}
c^{\dagger}_{\bm{k}\sigma} (\hat{h}_{\bm{k}})_{\sigma\sigma'}
c_{\bm{k}\sigma'} ,\\
\hat{h}_{\bm{k}}&=\xi_{\bm k} \hat{I}+\alpha(k_y\hat{\sigma}_x-k_x\hat{\sigma}_y)+\beta(k_x\hat{\sigma}_x-k_y\hat{\sigma}_y).
\end{align}
Here, $c_{{\bm k}\sigma}$ is an annihilation operator of conduction electrons with a wavenumber ${\bm k}=(k_x,k_y)$ and a spin $\sigma$ ($=\uparrow,\downarrow$), $\xi_{\bm k} = \hbar^2{\bm k}^2/2 m^*-\mu$ is the kinetic energy measured from the chemical potential $\mu$, $m^*$ is the effective mass, $\hat{I}$ is a $2\times 2$ identity matrix, and ${\bm \sigma} = (\sigma_x,\sigma_y,\sigma_z)$ denotes the Pauli matrices. The $2\times 2$ matrix $\hat{h}_{\bm{k}}$ includes the Rashba and Dresselhaus spin-orbit interactions, with strengths denoted by $\alpha$ and $\beta$, respectively.

\begin{figure}[tb]
\centering
\includegraphics[width=160mm]{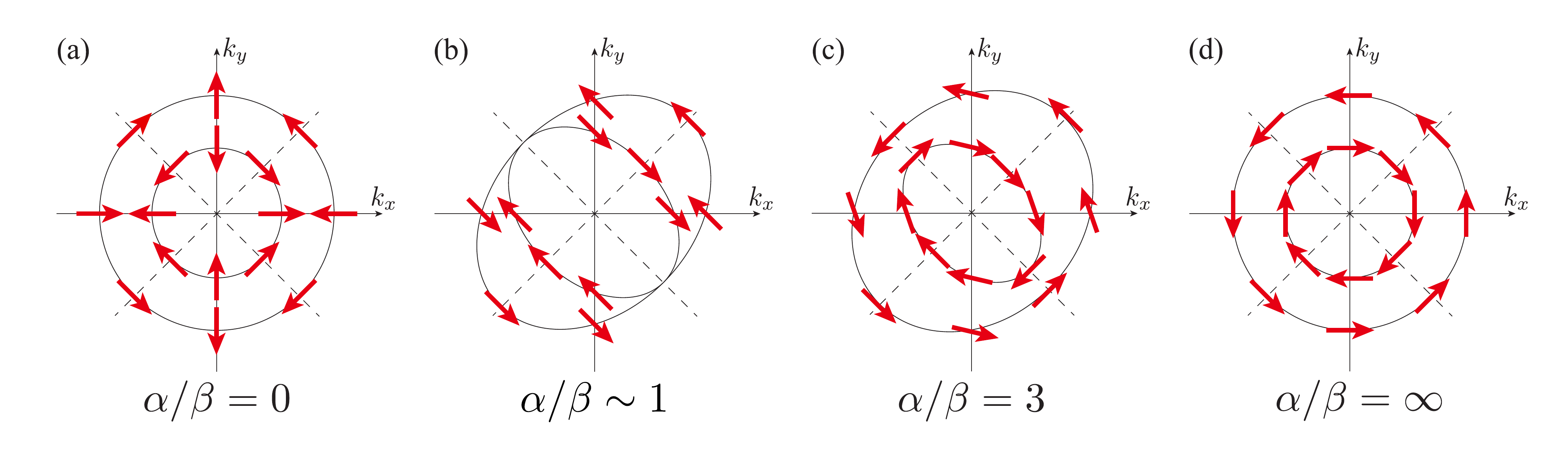}
\caption{The schematic picture of the spin-split Fermi surface for (a) $\alpha/\beta=0$, (b) $\alpha/\beta\sim 1$, (c) $\alpha/\beta=3$, and (d) $\alpha/\beta=\infty$. 
Red arrows represent spin polarization of 2DEG electrons in each Fermi surface.}
\label{fig:FS}
\end{figure}

In the presence of spin-orbit interaction, the Fermi surface is separated into two spin-polarized Fermi surfaces.
The quantization axis of the spin can be determined by an effective Zeeman field ${\bm h}_{\rm eff}({\bm k})$ acting on the conduction electrons, which is defined as
\begin{align}
\hat{h}_{\bm k} &=\xi_{\bm k} \hat{I} - {\bm h}_{\rm eff}({\bm k}) \cdot {\bm \sigma}, \\
{\bm h}_{\rm eff}({\bm k}) &=k\left(\begin{array}{c} -\alpha\sin\varphi -\beta\cos\varphi \\
\alpha\cos\varphi +\beta\sin\varphi \\ 0 \end{array} \right) ,
\label{eq:heffkdep}
\end{align}
where we have introduced the polar representation ${\bm k}=(k\cos \varphi,k\sin \varphi)$ for the electron wavenumber.
The quantization axis of the electron spin depends on the azimuth angle of the electron wavenumber, and its dependence changes by controlling the ratio between the two types of spin-orbit interaction.
The spin polarization near the Fermi surface is schematically shown for $\alpha/\beta = 0, 1, 3$ and $\infty$ in Fig.~\ref{fig:FS}.
For 2DEG realized in semiconductor heterostructures, this ratio can be tuned by sample fabrication and an external electric field.
The present model can also be considered as a prototype of systems that have a more complex electronic structure with spin-splitting bands.
We also consider impurities in 2DEG with a short-range potential $v({\bm r}) = v_0 \delta({\bm r})$, whose Hamiltonian is given as
\begin{align}
 {\cal H}_{\rm imp} &= \frac{v_0}{{\cal A}}\sum_{{\bm k},{\bm q},\sigma} \rho_{\rm imp}({\bm q})c^\dagger_{{\bm k}+{\bm q}\sigma} c_{{\bm k}\sigma} , 
 \label{eq:Himp2d}
\end{align}
where ${\cal A}$ is an area of 2DEG,
$\rho_{\rm imp}({\bm q}) = \sum_i e^{-i{\bm q}\cdot {\bm R}_i}$ and ${\bm R}_i$ denotes the position of the impurity.

The uniform spin precession of the FI is described by the following Hamiltonian:
\begin{align}
{\cal H}_{\rm FI} &=\hbar \omega_{\bm 0} b_{\bm 0}^{\dagger} b_{\bm 0},
\label{HamFI}
\end{align}
where $b_{\bm 0}$ is an annihilation operator of a uniform magnon mode, and $\omega_{\bm 0}$ is a ferromagnetic resonance (FMR) frequency.
Although an external microwave is described by a time-dependent magnetic field acting on the uniform magnon mode, we treat it in an indirect way as described in the subsequent sections.

Finally, let us introduce an interfacial coupling between the FI and 2DEG. 
We assume a clean interface at which the in-plane momentum conservation law holds for conduction electrons of 2DEG. 
The Hamiltonian for interfacial exchange interaction between the uniform magnon mode and 2DEG is given as
\begin{align}
{\cal H}_{\rm int} &= \sqrt{2S_0}(\mathcal{T} b_{\bm{0}} s^{x'-}_{\bm{0}}+\mathcal{T}^* b^\dagger_{\bm{0}} s^{x'+}_{\bm{0}}), \label{eq:Hint2}
\end{align}
where $\mathcal{T}$ denotes a strength of the interfacial exchange coupling.
The factor $\sqrt{2S_0}$ comes from the spin-wave approximation, where $S_0$ is an amplitude of the localized spin in the FI.
Here, we have introduced a new coordinate $(x',y',z')$ fixed to the spin polarization in the FI as shown in Fig.~\ref{fig:setup}~(b).
The spin ladder operator $s^{x'\pm}_{\bm{0}}=s_{\pm \bm q}^{y'}\pm i s_{\pm \bm q}^{z'}$ is defined in this new coordinate as
\begin{align}
s_{\bm q}^{x'} &= \cos \theta s_{\bm q}^{x} + \sin \theta s_{\bm q}^{y}, \\
s_{\bm q}^{y'} &= -\sin \theta s_{\bm q}^{x} + \cos \theta s_{\bm q}^{y}, \\
s_{\bm q}^{z'} &= s_{\bm q}^{z}, \\
s_{\bm q}^{a} &= \frac12 \sum_{\sigma\sigma'}
\sum_{{\bm k}} c_{{\bm k}\sigma}^\dagger (\sigma_a)_{\sigma \sigma'}
c_{{\bm k}+{\bm q}\sigma'}, \quad (a=x,y,z).
\end{align}
In the following calculation, we consider second-order perturbation with respect to ${\cal H}_{\rm int}$, assuming that the spin-splitting energy is much smaller than the Fermi energy and much larger than impurity scattering rate and temperature.

\section{Spin pumping}
\label{sec:title}

In this section, we study the increase of the Gilbert damping in the FMR measurement, which is one of the main targets in spin pump experiments. 
For a detailed calculation, refer to Refs.~\citeonline{Yama2021,Yama2023a}.

\subsection{Formulation}
\label{sec:SPformulation}

The damping rate in the FMR measurement can be obtained from the spin correlation function.
Let us start with an imaginary-time spin correlation function for an isolated FI:
\begin{align}
G_0(i\omega_n) 
&= -\frac{2S_0}{\hbar} \int_0^{\hbar/k_{\rm B}T} d\tau \, e^{i\omega_n \tau} \langle 
    b_{{\bm 0}}(\tau) b_{{\bm 0}}^\dagger(0)
    \rangle,
\end{align}
where $\omega_n = 2nk_{\rm B}T\pi/\hbar$ is the bosonic Matsubara frequency, $T$ is the temperature, $b_{\bm 0}(\tau) = e^{\tau/k_{\bm B}T} b_{\bm 0}e^{-\tau/k_{\bm B}T}$, and the factor $2S_0$ appears in transforming the spin operator into the magnon operator in the spin-wave approximation.
The spin correlation function is calculated from the Hamiltonian, Eq.~(\ref{HamFI}), as
\begin{align}
G_0(i\omega_n) & =\frac{2S_0/\hbar}{i\omega_n-\omega_{\bm 0}-\alpha_{\rm G}|\omega_n|}, \label{eq:correlation_magnon}
\end{align}
where we introduce $\alpha_{\rm G}$ as a phenomenological dimensionless parameter that represent the Gilbert damping in the bulk FI due to spin-phonon coupling.
Using the linear response theory, we can show that the microwave absorption rate is described by ${\rm Im}\, G_0^R(\omega)$, where $G_0^R(\omega)=G_0(i\omega_n\rightarrow \omega+i\delta)$ is a retarded component of the spin correlation function.\cite{Funato2022}

When the interfacial exchange coupling is introduced, the second-order perturbation with respect to ${\cal H}_{\rm int}$ gives
\begin{align}
G(i\omega_{n}) &= \frac{1}{(G_0(i\omega_n))^{-1}-\Sigma(i\omega_n)}, \\
\Sigma(i\omega_n) 
&=|\mathcal{T}|^2 {\cal A} \chi(i\omega_n), \label{eq:SelfEnergyChi}
\end{align}
where $\Sigma(i\omega_n)$ is the self-energy due to the interfacial exchange coupling and $\chi(i\omega_n)$ is the uniform spin susceptibility for conduction electrons per unit area, defined as
\begin{align}
\chi(i\omega_n) &= - \frac{1}{\hbar {\cal A}}\int_0^{\hbar/k_{\rm B}T} d\tau \, e^{i\omega_n \tau}\langle s_{\bm 0}^{x'+}(\tau) s_{\bm 0}^{x'-}(0) \rangle, 
\end{align}
where $s_{\bm 0}^{x'\pm}(\tau)=e^{{\cal H}_{\rm 2DEG}\tau/\hbar}s_{\bm 0}^{x'\pm} e^{-{\cal H}_{\rm 2DEG}\tau/\hbar}$ and ${\cal H}_{\rm 2DEG} = {\cal H}_{\rm kin} + {\cal H}_{\rm imp}$.
The retarded spin correlation function is obtained by analytic continuation $i\omega_n \rightarrow \omega + i\delta$, as
\begin{align}
G^R(\omega) 
&= \frac{2S_{0}/\hbar}{\omega-(\omega_{\bm 0}+\delta \omega_{\bm 0}) + i(\alpha_{\rm G}+\delta \alpha_{\rm G}) \omega},\label{eq:Green's_magnon} \\
\frac{\delta \omega_{\bm 0}}{\omega_{\bm 0}} &\simeq \frac{2S_0|\mathcal{T}|^2 {\cal A} }{\hbar\omega_{\bm 0}}\, {\rm Re}\, \chi^R(\omega_{\bm 0}),
\label{eq:deltaomega} \\
\delta \alpha_{\rm G} &\simeq -\frac{2S_{0}|\mathcal{T}|^2 {\cal A}}{\hbar\omega_{\bm 0}}\, {\rm Im}\, \chi^R(\omega_{\bm 0}),
\label{eq:deltagamma}
\end{align}
where $\delta \omega_{\bm 0}$ and $\delta\alpha_{\rm G}$ are respectively the changes in the FMR frequency and Gilbert damping due to the FI/2DEG interface, and $\chi^R(\omega)=\chi(i\omega_n\rightarrow \omega+i\delta)$. 
Thus, both the FMR frequency shift and the modulation of the Gilbert damping are determined by the uniform spin susceptibility of the conduction electrons, $\chi(i\omega_n)$.
We should note that our formulation does not need to use ``spin current'', for which technical problems due to the breaking of the spin conservation in the presence of the spin-orbit interaction have been discussed.\cite{Khaetskii2006, Shitade2022}

\subsection{Effect of vertex corrections}

\begin{figure}[tb]
\centering
\includegraphics[width=160mm]{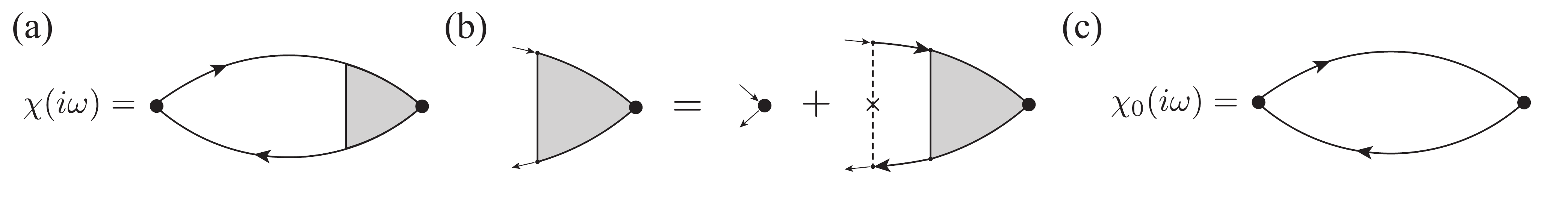}
\caption{The Feynman diagrams related to the calculation of $\chi(i\omega)$: (a) The diagram with the vertex function (represented by a gray hatch), (b) the diagrammatic expression of the (approximate) Bethe–Salpeter equation for the vertex function, and (c) the diagram without the vertex correction. 
The solid line represents the propagator of conduction electrons with the impurity self-energy in the Born approximation, while the vertical dashed line with a cross represents the impurity scattering.}
\label{fig:Diagram}
\end{figure}

From now on, we briefly explain how to calculate $\chi(i\omega)$ using the standard diagrammatic method (for a detailed calculation, see Refs.~\citeonline{Yama2021,Yama2023a}).
We calculate the electron propagator using an impurity self-energy by the Born approximation.
The Feynman diagram for $\chi(i\omega)$ is shown in Fig.~\ref{fig:Diagram}(a), where the solid line indicates the electron propagator with the impurity self-energy, and the gray triangle represents the vertex function.
The vertex function is determined by the Bethe-Salpeter equation constructed from the inpurity self-energy through the Ward identity and is schematically shown in Fig.~\ref{fig:Diagram}(b) for the Born approximation.
We stress that the vertex function plays an important role in constructing the self-consistent approximation, which satisfies various conservation laws.
For comparison, let us consider a simpler spin correlation function $\chi_0(i\omega)$, whose diagram is given in Fig.~\ref{fig:Diagram}(c).
The difference, $\chi(i\omega_n)-\chi_0(i\omega_n)$, is referred to as the vertex correction.

\begin{figure}[tb]
\centering
\includegraphics[width=140mm]{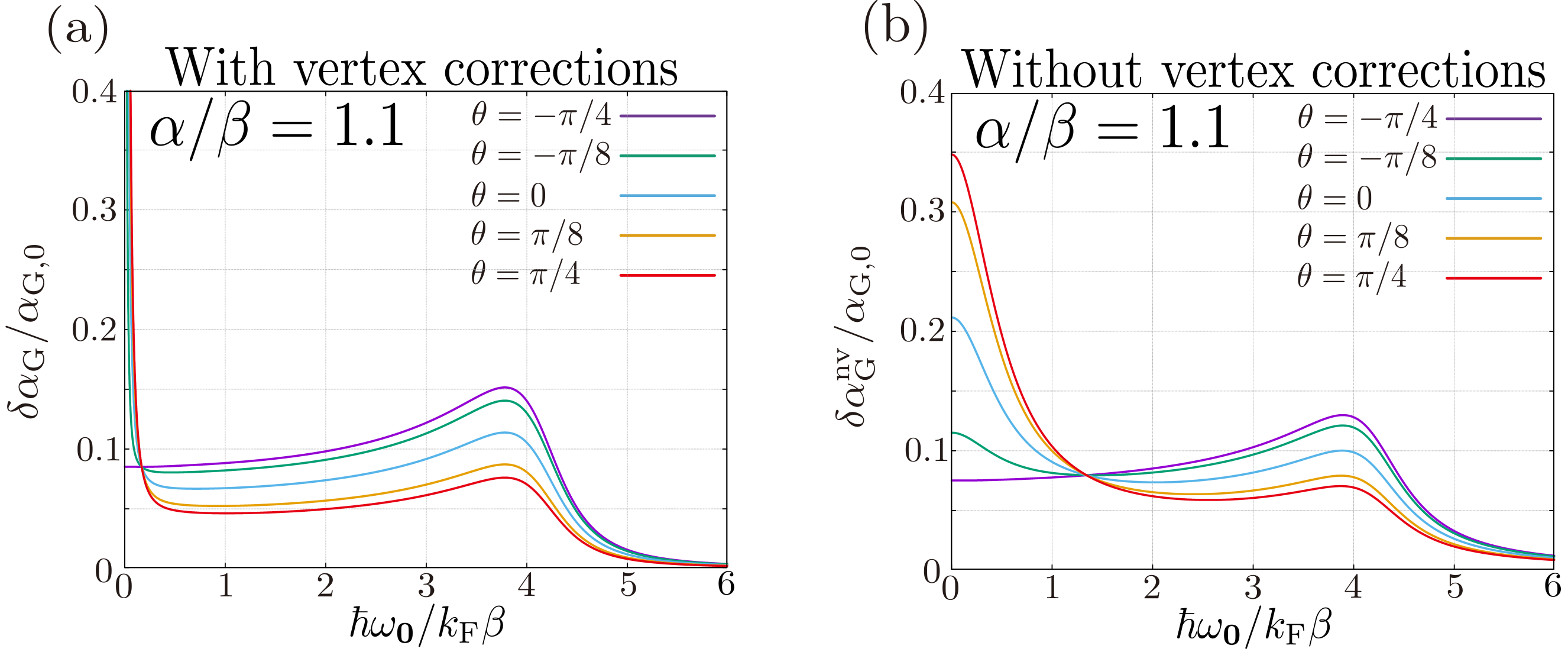}
\caption{Modulation of the Gilbert damping (a) with vertex corrections, $\delta \alpha_{\rm G}$ and (b) without vertex corrections, $\delta \alpha_{{\rm G}}^{\rm nv}$, is plotted as a function of the FMR frequency $\omega_{\bm 0}$. The energy broadening due to the impurities is taken as $\Gamma = k_{\rm B}\beta$.
The constant parameter $\alpha_{{\rm G},0}$ includes only the information of the interfacial coupling and is independent of $\theta$ and $\omega_{\bm 0}$.}
\label{fig:GilbertDamping}
\end{figure}

Fig.~\ref{fig:GilbertDamping}~(a) shows the result for the increase of the Gilbert damping, $\delta\alpha_{\rm G}$, for $\alpha/\beta = 1.1$ taking into account the vertex corrections.
The case of $\alpha/\beta\simeq 1$ is special because the effective Zeeman field ${\bm h}_{\rm eff}$ points almost in the $\pi/4$ direction (see also Fig.~\ref{fig:FS}(b)). 
A broad peak in the range of $0\le \hbar \omega_{\bm 0} \le k_{\rm F}\beta$ corresponds to the distribution of the spin splitting energy $|{\bm h}_{\rm eff}|$ along the Fermi surface.
In addition, a sharp peak appears near the zero frequency.
The width of this sharp peak approaches zero in the limit of $\alpha/\beta \rightarrow 1$ (not shown in the figure).
This result indicates that the sharp peak at $\omega_{\bm 0} = 0$ is related to an inverse of the spin relaxation time of 2DEG, which becomes infinite just at $\alpha/\beta =1$.

Fig.~\ref{fig:GilbertDamping}~(b) shows the result for the increase of the Gilbert damping, $\delta\alpha^{\rm nv}_{\rm G}$, without vertex corrections, i.e., the one calculated from $\chi_0(i\omega)$.
Compared with Fig.~\ref{fig:GilbertDamping}~(a), 
the broad peak structure is reproduced, while the peak at $\omega_{\bm 0} = 0$ is largely broadened.
This result clarifies the importance of the vertex corrections, which incorporate the spin conservation law.

\section{INVERSE RASHBA-EDELSTEIN EFFECT}

Next, we study the current induced by the inverse Rashba-Edelstein effect in 2DEG, which is driven by spin pumping.
For a detailed calculation, refer to Ref.~\citeonline{Yama2023b}.

\subsection{Formulation}

In our study, we are focusing on the case that the spin-splitting energy in 2DEG is much larger than the energy broadening due to the impurity scattering, while it is much smaller than the Fermi energy. 
In this condition, the distribution function of conduction electrons in 2DEG can be expressed for a uniform steady state as $f({\bm k},\gamma)$, where $\gamma$ is an index of the spin-polarized bands\cite{Suzuki2023}.
The Boltzmann equation for our model is described as
\begin{align}
0 = \frac{\partial f}{\partial t}\biggl{|}_{\rm pump}+\frac{\partial f}{\partial t}\biggl{|}_{\rm imp} , \label{eq:0ecoll}
\end{align}
where $\partial f/\partial t|_{\rm pump}$ is a collision term due to spin injection from the FI into the 2DEG through the interface, and $\partial f/\partial t|_{\rm imp}$ is a collision term due to impurity scattering. 
For the linear response to external driving, the distribution function can be approximated as\cite{Ziman1960, Lundstrom2000}
\begin{align}
f(\bm{k},\gamma) \simeq f_0(E^{\gamma}_{\bm{k}})-\frac{\partial f_0(E^{\gamma}_{\bm{k}})}{\partial E^{\gamma}_{\bm{k}}} \delta\mu(\varphi,\gamma).
\label{eq:fexpan}
\end{align}
where $E^{\gamma}_{\bm{k}}$ is an energy dispersion of electrons in 2DEG, $f_0(\epsilon) =(\exp[ (\epsilon-\mu)k_{\rm B}T]+1)^{-1}$ is the Fermi distribution function, and $\delta \mu(\varphi,\gamma)$ is a chemical potential shift in the direction of $\varphi$.

Spin injection from the FI into the 2DEG is described by stochastic excitation induced by magnon absorption and emission. This process can be expressed by the collision term as
\begin{align}
& \frac{\partial f(\bm{k},\gamma)}{\partial t}\biggl{|}_{\rm pump} 
=\sum_{\bm{k}'}\sum_{\gamma'=\pm}
\Bigl{[}
P_{\bm{k}'\gamma'\rightarrow\bm{k}\gamma}
f(\bm{k}',\gamma')(1-f(\bm{k},\gamma))
-P_{\bm{k}\gamma\rightarrow\bm{k}'\gamma'}
f(\bm{k},\gamma)(1-f(\bm{k}',\gamma'))
\Bigl{]}, \label{eq:colpump1}
\end{align}
where $P_{\bm{k}\gamma\rightarrow\bm{k}'\gamma'}$ is the transition rate calculated with Fermi’s golden rule as
\begin{align}
P_{\bm{k}\gamma\rightarrow\bm{k}'\gamma'}
=\sum_{N_{\bm{0}}}\sum_{\Delta N_{\bm{0}} = \pm 1}
\frac{2\pi}{\hbar}
\Bigl{|}\langle \bm{k}',\gamma'|\langle N_{\bm{0}}+\Delta N_{\bm{0}} |{\cal H}_{\rm int}
|\bm{k},\gamma\rangle|N_{\bm{0}}\rangle \Bigl{|}^2\rho(N_{\bm{0}}) \delta\Bigl{(}E^{\gamma'}_{\bm{k}'}-E^{\gamma}_{\bm{k}}+\Delta N_{\bm{0}}\hbar\omega_{\bm{0}}\Bigl{)}, \label{eq:Pkkp}
\end{align}
where $|N_{\bm{0}}\rangle$ is the eigenstate of the magnon number operator, i.e., $b^{\dagger}_{\bm{0}}b_{\bm{0}}|N_{\bm{0}}\rangle=N_{\bm{0}}|N_{\bm{0}}\rangle$, $\Delta N_{\bm{0}}=\pm 1$ is a change of the magnon number, and $\rho(N_{\bm{0}})$ describes a nonequilibrium distribution function for the uniform spin precession driven by microwave. 
Assuming that the distribution function $\rho(N_{\bm{0}})$ has a sharp peak at its average $\langle N_{\bm{0}}\rangle$, the summation can then be approximated as $\sum_{N_{\bm{0}}} \rho(N_{\bm{0}}) F(N_{\bm{0}})
\simeq F(\langle N_{\bm{0}}\rangle)$, where $F(x)$ is an arbitrary function. 

The collision term due to impurity scattering is written as
\begin{align}
& \frac{\partial f(\bm{k},\gamma)}{\partial t}\biggl{|}_{\rm imp} =\sum_{\bm{k}'}\sum_{\gamma'=\pm}
\Bigl{[}
Q_{\bm{k}'\gamma'\rightarrow\bm{k}\gamma}
f(\bm{k}',\gamma')(1-f(\bm{k},\gamma))
-Q_{\bm{k}\gamma\rightarrow\bm{k}'\gamma'}
f(\bm{k},\gamma)(1-f(\bm{k}',\gamma'))
\Bigl{]}, \label{eq:colimp1}
\end{align}
where $Q_{\bm{k}\gamma\rightarrow\bm{k}'\gamma'}$ is the transition rate of electron scattering given as
\begin{align}
&Q_{\bm{k}\gamma\rightarrow\bm{k}'\gamma'} 
= \frac{2\pi}{\hbar}
\Bigl{|}\langle \bm{k}',\gamma'|H_{\rm imp}({\bm R})|\bm{k},\gamma\rangle \Bigl{|}^2 \delta\Bigl{(}E^{\gamma'}_{\bm{k}'}-E^{\gamma}_{\bm{k}}\Bigl{)}. \label{eq:Qkkp}
\end{align}
Note that the transition rates due to interfacial and impurity scattering include the overlap of the spin states between the initial and final states.

Combining these expressions of the collision terms with the Boltzmann equation, we can calculate the nonequilibrium distribution function $f({\bm k},\gamma)$ (for a detailed calculation, see Ref.~\citeonline{Yama2023b}).
Note note again that our formulation does not need spin current.

\subsection{Induced current}

\begin{figure}[tb]
\centering
\includegraphics[width=90mm]{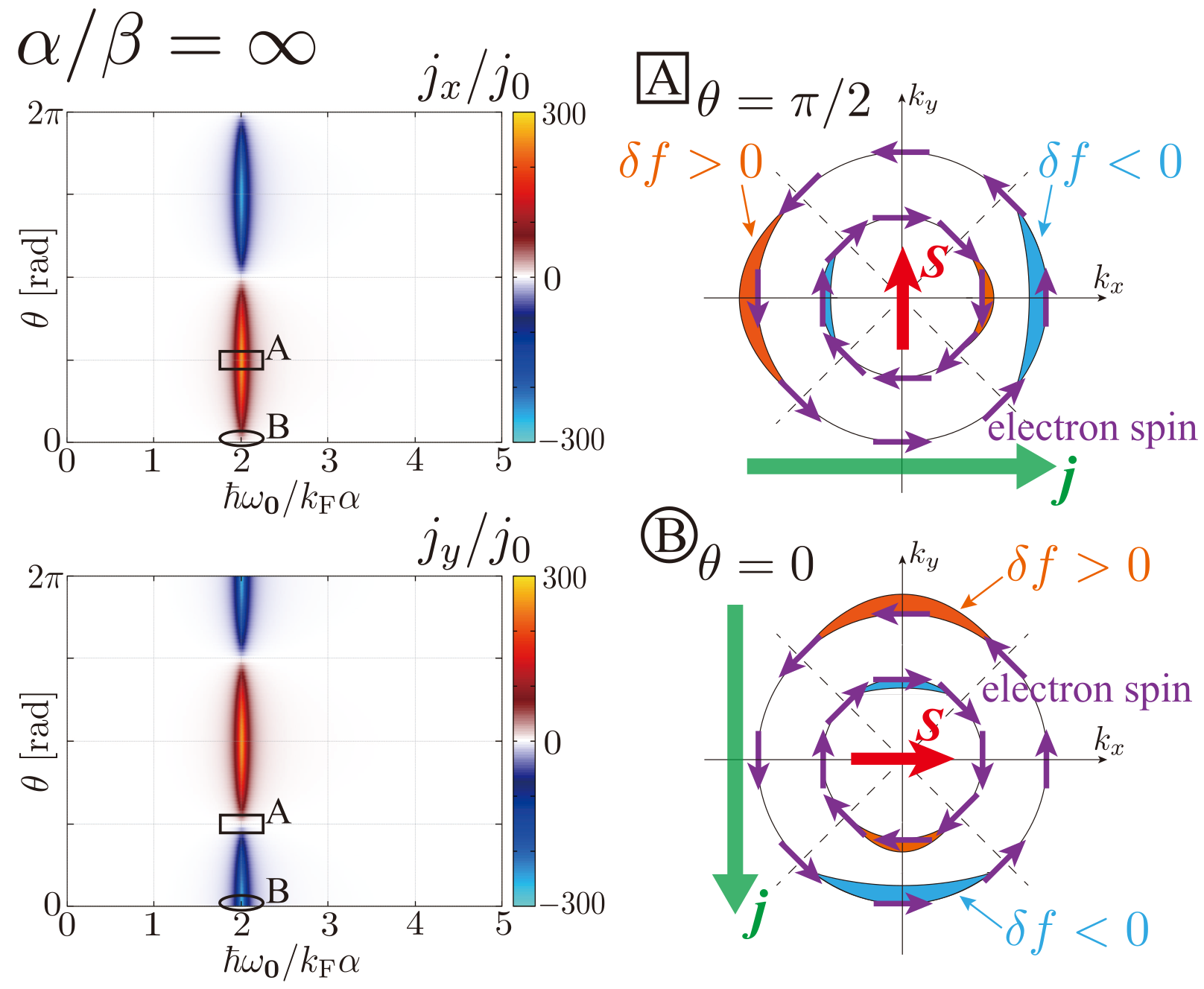}
\caption{Current density $\bm{j}=(j_x, j_y)$ for $\Gamma=0.1k_{\rm F}\alpha$ (left two panels) as a function of the FMR frequency $\omega_{\bm0}$ and the spin azimuth angle $\theta$ of the FI in the case that only the Rashba spin-orbit interaction exists ($\alpha/\beta=\infty$). 
The two right panels schematically show the change in the distribution function and the direction of the current for $\theta = \pi/2$ and $\theta = 0$, respectively, where the change in the distribution function is indicated by the orange and blue regions.} 
\label{fig:current1}
\end{figure}

From the distribution function $f({\bm k},\gamma)$, the current induced by spin pumping is calculated as
\begin{align}
\bm{j}&=\frac{e}{\mathcal{A}} \sum_{\bm{k},\gamma}\bm{v}(\bm{k},\gamma)f(\bm{k},\gamma),
\label{eq:IIREE}\\
\bm{v}(\bm{k},\gamma)&=
\frac{1}{\hbar}\frac{\partial E^{\gamma}_{\bm{k}}}{\partial\bm{k}}
=\frac{\hbar\bm{k}}{m^*}+\frac{\gamma}{\hbar}\frac{\partial h_{\rm eff}(\bm{k})}{\partial\bm{k}} ,
\end{align}
where $e$ ($<0$) is an electron charge, $\bm{v}(\bm{k},\gamma)$ is an electron velocity, and $h_{\rm eff}({\bm k})=|{\bm h}_{\rm eff}({\bm k})|$ is a spin-splitting energy.
The left two panels of Fig.~\ref{fig:current1} respectively show the result of $j_x$ and $j_y$ for $\alpha/\beta = \infty$, i.e., for the case where only the Rashba spin-orbit interaction exists.
These currents are shown by the contour plot as a function of the FMR frequency $\omega_{\bm 0}$ and the spin azimuth angle of the FI.
We find that a large current is induced when $\hbar \omega_{\bm 0}$ matches the spin-splitting energy $2k_{\rm F}\alpha$.

The direction of the induced current rotates when the spin azimuth angle $\theta$ changes. For $\theta=\pi/2$ (indicated by A in a square in the contour plots), the current density ${\bm j}$ is induced in the $+x$ direction. This result can be intuitively explained as follows. When the spin polarization of the FI relaxes towards the original state (the $+y$ direction when $\theta=\pi/2$), the spin component in the $+y$ direction transfers from the 2DEG to the FI. Consequently, the nonequilibrium distribution function decreases (increases) for the $+y$-spin ($-y$-spin) band. The regions of the Fermi surface where the distribution function increases (decreases) are schematically shown in orange (blue) in the right picture of Fig.~\ref{fig:current1}. Since the density of states on the outer Fermi surface is larger than that on the inner surface, this change in the distribution function results in a net flow of electrons in the $-x$ direction, inducing a current in the $+x$ direction. A similar explanation applies for $\theta=0$ (indicated by the ellipse B in the contour plots and the lower picture on the right).

\begin{figure}[tb]
\centering
\includegraphics[width=140mm]{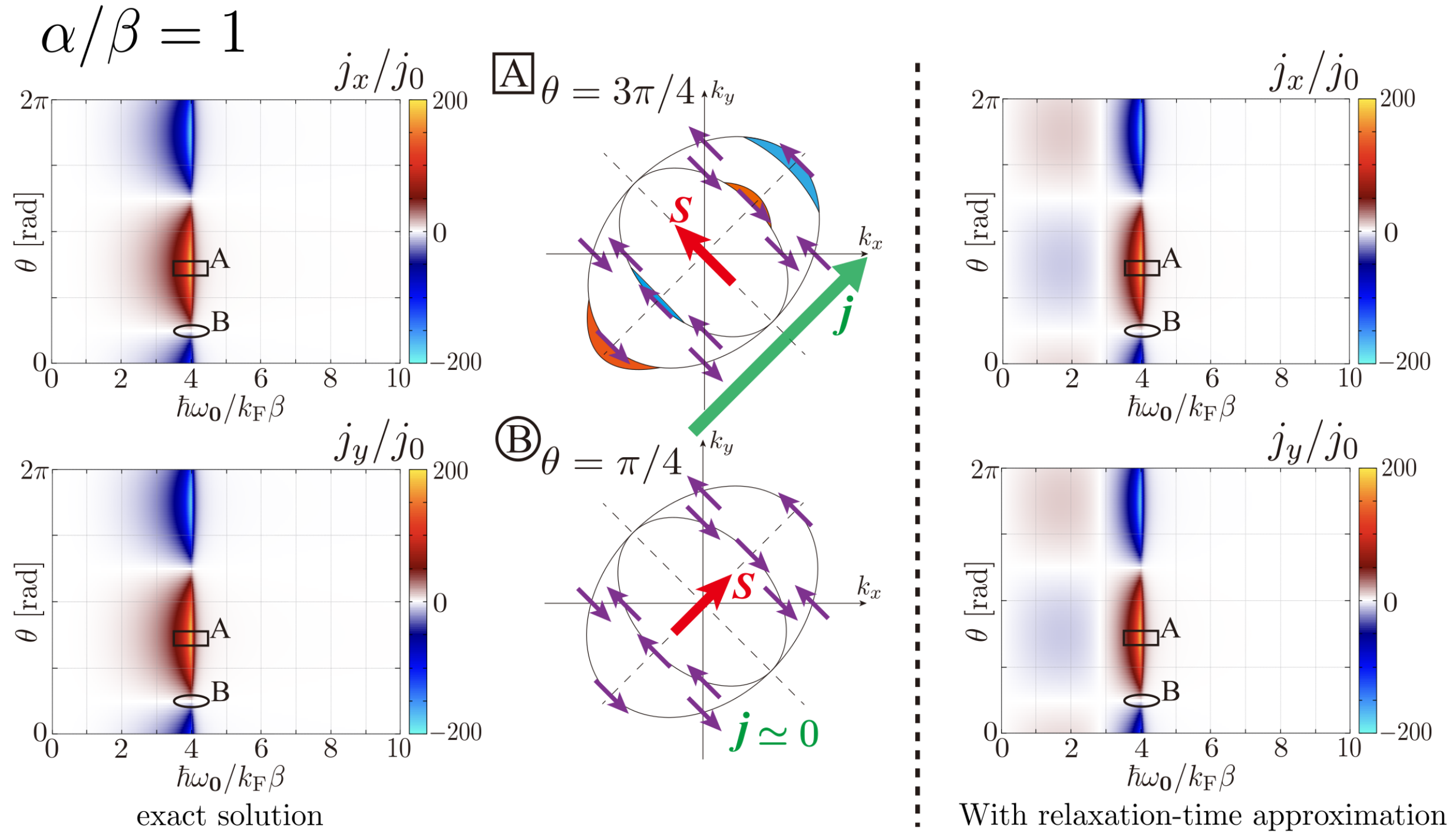}
\caption{Current density $\bm{j}=(j_x, j_y)$ for $\Gamma=0.1k_{\rm F}\alpha$ (left two panels) as a function of the FMR frequency $\omega_{\bm0}$ and the spin azimuth angle $\theta$ of the FI. 
The central two panels schematically show the change of the distribution function and the direction of the current for $\theta = 3\pi/4$ and $\theta = \pi/4$, respectively, where a change of the distribution function is indicated by the orange and blue regions.} 
\label{fig:current2}
\end{figure}

Next, we consider the case of $\alpha/\beta=1$, where the Dresselhaus and Rashba spin-orbit interactions compete with each other. The two plots on the left of Fig.~\ref{fig:current2} show the current density as functions of $\omega_{\bm{0}}$ and $\theta$. 
Due to the distribution of $h_{\rm eff}(\varphi)$, the current densities exhibit large values across a wide range of $0 \lesssim \hbar\omega_{\bm{0}} \lesssim 4 k_{\rm F}\beta$. 
The amplitude of ${\bm j}$ takes a maximum at $\theta=3\pi/4$ or $7\pi/4$ and almost vanish at $\theta=\pi/4$ or $5\pi/4$.
This distinctive result can be explained as follows (see the central two schematic pictures of Fig.~\ref{fig:current2}). 
For $\theta =3\pi/4$ (indicated by A in the contour plots in Fig.~\ref{fig:current2}), the distribution function changes in the direction of $\varphi = \pi/4$ and $5\pi/4$.
This change in the distribution function leads to a net electron flow (current) in the direction of $\varphi = 5\pi/4$ ($\varphi = \pi/4$). In contrast, for $\theta =\pi/4$ (indicated by B in the contour plots in Fig.~\ref{fig:current2}), no current flows.

In the calculation of the induced current, it is important to solve the Boltzmann equation without any approximation.
For comparison, we consider the relaxation-time approximation\cite{Silsbee2004}
\begin{align}
\left.\frac{\partial f({\bm k},\gamma)}{\partial t}\right|_{\rm imp} &= -\frac{f({\bm k},\gamma)-f_0({\bm k},\gamma)}{\tau({\bm k},\gamma)}, \\
\tau({\bm k},\gamma) &= \tau (1-\gamma \zeta(\varphi)),\\
\zeta(\varphi)&\equiv
\frac{2\pi D(\epsilon_{\rm F})}{k_{\rm F}}\sqrt{\alpha^2+\beta^2+2\alpha\beta\sin2\varphi},
\end{align}
where $\tau$ is a spin-independent relaxation time, $\zeta(\varphi)$ is a modulation of it due to the spin-orbit interaction, and $D(\epsilon_{\rm F})$ is the density of states near the Fermi energy.
The two plots on the right of Fig.~\ref{fig:current2} show the current densities calculated by this approximation.
We find that in this approximation the current changes sign for $\hbar\omega_{\bm 0} \lesssim 3 k_{\rm F}\beta$, indicating a significant difference from the full solution of the Boltzmann equation.

\subsection{Effect of spin relaxation} 

\begin{figure}[tb]
\centering
\includegraphics[width=120mm]{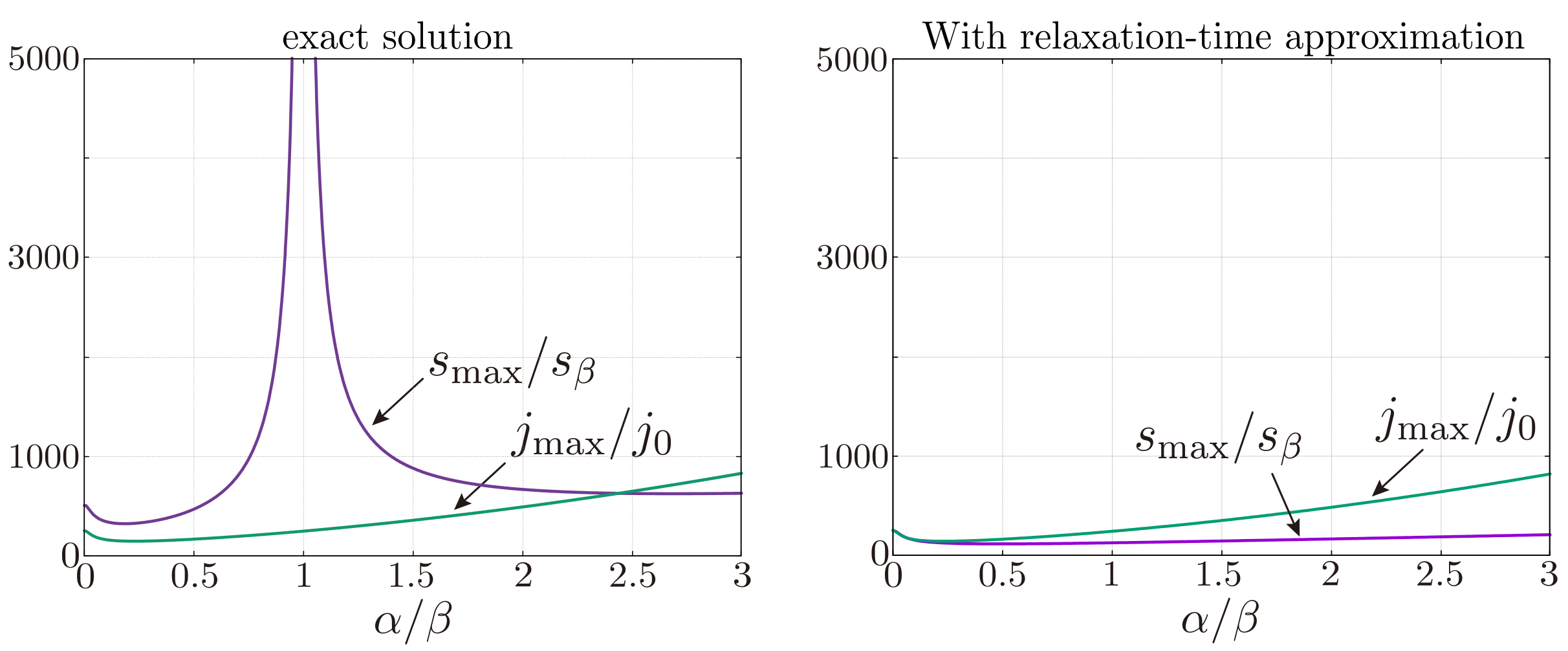}
\caption{Left panel: maximum spin density $s_{\rm max}\equiv\max(\sqrt{s_x^2+s_y^2})$ and maximum current density $j_{\rm max}\equiv\max(\sqrt{j_x^2+j_y^2})$ are plotted as a function of $\alpha/\beta$ for $\Gamma/k_{\rm F}\beta$=0.1.
Right panel: the results obtained by the relaxation-time approximation.
}
\label{fig:max}
\end{figure}

Finally, we discuss why the relaxation-time approximation leads to inappropriate results in more detail. Fig.~\ref{fig:max} shows the maximum spin density $s_{\rm max}\equiv\max(\sqrt{s_x^2+s_y^2})$ and the maximum current density $j_{\rm max}\equiv\max(\sqrt{j_x^2+j_y^2})$ of the 2DEG as functions of $\alpha/\beta$. The left and right panels of Fig.\ref{fig:max} display the results obtained by the full solution of the Boltzmann equation and the relaxation-time approximation, respectively.
We observe that the spin accumulation diverges at $\alpha/\beta = 1$ in the full solution of the Boltzmann equation (left panel). This reflects the special property of the 2DEG at $\alpha = \beta$, where the direction of the spin quantization axis remains unchanged along the Fermi surface (see Fig.~\ref{fig:FS}(b)). At this point, the spin component in the direction of $3\pi/4$ is conserved, resulting in a divergent spin relaxation time and, consequently, divergent spin accumulation at $\alpha = \beta$.
In contrast, the relaxation-time approximation (right panel) fails to reproduce this divergent behavior at $\alpha = \beta$. 

It is worth noting that the full solution of the Boltzmann equation corresponds to full consideration of the vertex corrections in the linear response theory\cite{Datta1997,Inoue2004,Inoue2006}.
Therefore, the inaccurate description of spin accumulation and current in the relaxation-time approximation reflects the imperfect incorporation of the conservation law in 2DEG.

\section{SUMMARY}

We theoretically investigated spin pumping into a two-dimensional electron gas (2DEG) with Rashba- and Dresselhaus-type spin-orbit interactions, which is regarded as a prototype system. We formulated the change in the peak position and the linewidth in ferromagnetic resonance (FMR) in terms of the spin correlation function in 2DEG. Additionally, we developed a formulation for the current density induced by the Rashba-Edelstein effect in 2DEG based on the Boltzmann equation. We emphasized that the conservation laws incorporated in the vertex corrections or the full solution of the Boltzmann equation play an important role in accurately describing these physical phenomena. Our formulation and results, which do not use the concept of spin current, will be helpful for describing spin pumping and the resulting current generation in 2DEG realized in surface/interface states and atomic layer compounds.

\acknowledgments 
 
The authors thank Y. Suzuki, Y. Kato, and M. Kohda for their helpful discussions. 
M. Y. was supported by JST SPRING (Grant No.~JPMJSP2108) and JSPS KAKENHI Grant Number JP24KJ0624. M. M. was supported by the National Natural Science Foundation of China (NSFC) under Grant No. 12374126, by the Priority Program of Chinese Academy of Sciences under Grant No.~XDB28000000, and by JSPS KAKENHI for Grants (No.~JP21H04565, No.~JP21H01800, No.~JP23H01839, and No.~24H00322) from MEXT, Japan.
T.K. was supported by JSPS KAKENHI Grant No.~JP24K06951.

\bibliography{report} 

\begin{thebibliography}{10}

\bibitem{Tserkovnyak2002}
Tserkovnyak, Y., Brataas, A., and Bauer, G. E.~W., ``Enhanced gilbert damping in thin ferromagnetic films,'' {\em Phys. Rev. Lett.}~{\bf 88},  117601 (2002).

\bibitem{Tserkovnyak2005}
Tserkovnyak, Y., Brataas, A., Bauer, G. E.~W., and Halperin, B.~I., ``Nonlocal magnetization dynamics in ferromagnetic heterostructures,'' {\em Rev. Mod. Phys.}~{\bf 77},  1375--1421 (2005).

\bibitem{Mizukami2001}
Mizukami, S., Ando, Y., and Miyazaki, T., ``{The study on ferromagnetic resonance linewidth for NM/80NiFe/NM (NM= Cu, Ta, Pd and Pt) films},'' {\em Jpn. J. Appl. Phys.}~{\bf 40}(2R),  580 (2001).

\bibitem{Saitoh2006}
Saitoh, E., Ueda, M., Miyajima, H., and Tatara, G., ``{Conversion of spin current into charge current at room temperature: Inverse spin-Hall effect},'' {\em Appl. Phys. Lett.}~{\bf 88},  182509 (2006).

\bibitem{Kajiwara2010}
Kajiwara, Y., Harii, K., Takahashi, S., Ohe, J., Uchida, K., Mizuguchi, M., Umezawa, H., Kawai, H., Ando, K., Takanashi, K., Maekawa, S., and Saitoh, E., ``{Transmission of electrical signals by spin-wave interconversion in a magnetic insulator},'' {\em Nature}~{\bf 464},  262 (2010).

\bibitem{Hellman2017}
Hellman, F., Hoffmann, A., Tserkovnyak, Y., Beach, G. S.~D., Fullerton, E.~E., Leighton, C., MacDonald, A.~H., Ralph, D.~C., Arena, D.~A., D\"urr, H.~A., Fischer, P., Grollier, J., Heremans, J.~P., Jungwirth, T., Kimel, A.~V., Koopmans, B., Krivorotov, I.~N., May, S.~J., Petford-Long, A.~K., Rondinelli, J.~M., Samarth, N., Schuller, I.~K., Slavin, A.~N., Stiles, M.~D., Tchernyshyov, O., Thiaville, A., and Zink, B.~L., ``Interface-induced phenomena in magnetism,'' {\em Rev. Mod. Phys.}~{\bf 89},  025006 (2017).

\bibitem{Han2020}
Han, W., Maekawa, S., and Xie, X.-C., ``Spin current as a probe of quantum materials,'' {\em Nat. Mater.}~{\bf 19},  139 (2020).

\bibitem{Hirobe2019}
Hirobe, D., Sato, M., Hagihala, M., Shiomi, Y., Masuda, T., and Saitoh, E., ``Magnon pairs and spin-nematic correlation in the spin seebeck effect,'' {\em Phys. Rev. Lett.}~{\bf 123},  117202 (2019).

\bibitem{Kato2019}
Kato, T., Ohnuma, Y., Matsuo, M., Rech, J., Jonckheere, T., and Martin, T., ``Microscopic theory of spin transport at the interface between a superconductor and a ferromagnetic insulator,'' {\em Phys. Rev. B}~{\bf 99},  144411 (2019).

\bibitem{Yamamoto2021}
Yamamoto, T., Kato, T., and Matsuo, M., ``{Spin current at a magnetic junction as a probe of the Kondo state},'' {\em Phys. Rev. B}~{\bf 104},  L121401 (2021).

\bibitem{Ominato2022a}
Ominato, Y., Yamakage, A., Kato, T., and Matsuo, M., ``Ferromagnetic resonance modulation in $d$-wave superconductor/ferromagnetic insulator bilayer systems,'' {\em Phys. Rev. B}~{\bf 105},  205406 (2022).

\bibitem{Ominato2022b}
Ominato, Y., Yamakage, A., and Matsuo, M., ``Anisotropic superconducting spin transport at magnetic interfaces,'' {\em Phys. Rev. B}~{\bf 106},  L161406 (2022).

\bibitem{Funato2022}
Funato, T., Kato, T., and Matsuo, M., ``Spin pumping into anisotropic dirac electrons,'' {\em Phys. Rev. B}~{\bf 106},  144418 (2022).

\bibitem{Fukuzawa2023}
Fukuzawa, K., Kato, T., Matsuo, M., Jonckheere, T., Rech, J., and Martin, T., ``Spin pumping into carbon nanotubes,'' {\em Phys. Rev. B}~{\bf 108},  134429 (2023).

\bibitem{Haddad2023}
Haddad, S., Kato, T., Zhu, J., and Mandhour, L., ``Twisted bilayer graphene reveals its flat bands under spin pumping,'' {\em Phys. Rev. B}~{\bf 108},  L121101 (2023).

\bibitem{Shen2014a}
Shen, K., Vignale, G., and Raimondi, R., ``{Microscopic Theory of the Inverse Edelstein Effect},'' {\em Phys. Rev. Lett.}~{\bf 112},  096601 (2014).

\bibitem{Manchon2015}
Manchon, A., Koo, H.~C., Nitta, J., Frolov, S.~M., and Duine, R.~A., ``{New perspectives for Rashba spin-orbit coupling},'' {\em Nat. Mater.}~{\bf 14},  871--882 (2015).

\bibitem{Soumyanarayanan2016}
Soumyanarayanan, A., Reyren, N., Fert, A., and Panagopoulos, C., ``Emergent phenomena induced by spin--orbit coupling at surfaces and interfaces,'' {\em Nature}~{\bf 539}(7630),  509--517 (2016).

\bibitem{Sanchez2013}
Rojas-S{\'a}nchez, J.-C.~R., Vila, L., Desfonds, G., Gambarelli, S., Attan{\'e}, J.~P., De~Teresa, J.~M., Mag{\'e}n, C., and Fert, A., ``Spin-to-charge conversion using rashba coupling at the interface between non-magnetic materials,'' {\em Nat. Commun.}~{\bf 4},  2944 (2013).

\bibitem{Lesne2016}
Lesne, E., Fu, Y., Oyarzun, S., Rojas-S{\'a}nchez, J.~C., Vaz, D.~C., Naganuma, H., Sicoli, G., Attan{\'e}, J.-P., Jamet, M., Jacquet, E., George, J.-M., Barth\'{e}l\'{e}my, A., Jaffr\`{e}s, H., Fert, A., Bibes, M., and Vila, L., ``{Highly efficient and tunable spin-to-charge conversion through Rashba coupling at oxide interfaces},'' {\em Nat. Mater.}~{\bf 15},  1261 (2016).

\bibitem{Song2017}
Song, Q., Zhang, H., Su, T., Yuan, W., Chen, Y., Xing, W., Shi, J., Sun, J., and Han, W., ``{Observation of inverse Edelstein effect in Rashba-split 2DEG between SrTiO${}_3$ and LaAlO${}_3$ at room temperature},'' {\em Sci. Adv.}~{\bf 3},  e1602312 (2017).

\bibitem{Bernevig2006}
Bernevig, B.~A., Orenstein, J., and Zhang, S.-C., ``Exact su(2) symmetry and persistent spin helix in a spin-orbit coupled system,'' {\em Phys. Rev. Lett.}~{\bf 97},  236601 (Dec 2006).

\bibitem{Kohda2017}
Kohda, M. and Salis, G., ``Physics and application of persistent spin helix state in semiconductor heterostructures,'' {\em Semicond. Sci. Technol.}~{\bf 32},  073002 (jun 2017).

\bibitem{Ganichev2004}
Ganichev, S.~D., Bel'kov, V.~V., Golub, L.~E., Ivchenko, E.~L., Schneider, P., Giglberger, S., Eroms, J., De~Boeck, J., Borghs, G., Wegscheider, W., Weiss, D., and Prettl, W., ``Experimental separation of rashba and dresselhaus spin splittings in semiconductor quantum wells,'' {\em Phys. Rev. Lett.}~{\bf 92},  256601 (Jun 2004).

\bibitem{Trushin2007}
Trushin, M. and Schliemann, J., ``Anisotropic current-induced spin accumulation in the two-dimensional electron gas with spin-orbit coupling,'' {\em Phys. Rev. B}~{\bf 75},  155323 (Apr 2007).

\bibitem{Yama2021}
Yama, M., Tatsuno, M., Kato, T., and Matsuo, M., ``Spin pumping of two-dimensional electron gas with rashba and dresselhaus spin-orbit interactions,'' {\em Phys. Rev. B}~{\bf 104},  054410 (2021).

\bibitem{Yama2023a}
Yama, M., Matsuo, M., and Kato, T., ``Effect of vertex corrections on the enhancement of gilbert damping in spin pumping into a two-dimensional electron gas,'' {\em Phys. Rev. B}~{\bf 107},  174414 (2023).

\bibitem{Khaetskii2006}
Khaetskii, A., ``{Nonexistence of Intrinsic Spin Currents},'' {\em Phys. Rev. Lett.}~{\bf 96},  056602 (2006).

\bibitem{Shitade2022}
Shitade, A. and Tatara, G., ``Spin accumulation without spin current,'' {\em Phys. Rev. B}~{\bf 105},  L201202 (2022).

\bibitem{Yama2023b}
Yama, M., Matsuo, M., and Kato, T., ``Theory of inverse rashba-edelstein effect induced by spin pumping into a two-dimensional electron gas,'' {\em Phys. Rev. B}~{\bf 108},  144430 (2023).

\bibitem{Suzuki2023}
Suzuki, Y. and Kato, Y., ``Spin relaxation, diffusion, and edelstein effect in chiral metal surface,'' {\em Phys. Rev. B}~{\bf 107},  115305 (2023).

\bibitem{Ziman1960}
Ziman, J.~M.,  [{\em Electrons and Phonons: The Theory of Transport Phenomena in Solids}{\nolinebreak\hspace{0.1em}]}, Clarendon Press, Oxford (1960).

\bibitem{Lundstrom2000}
Lundstrom, M.,  [{\em Fundamentals of Carrier Transport}{\nolinebreak\hspace{0.1em}]}, Cambridge University Press, Cambridge (2000).

\bibitem{Silsbee2004}
Silsbee, R.~H., ``Spin-orbit induced coupling of charge current and spin polarization,'' {\em J. Phys.: Condens. Matter}~{\bf 16},  R179 (2004).

\bibitem{Datta1997}
Datta, S.,  [{\em Electronic Transport in Mesoscopic Systems}{\nolinebreak\hspace{0.1em}]}, Cambridge Studies in Semiconductor Physics, Cambridge University Press (1997).

\bibitem{Inoue2004}
Inoue, J.-i., Bauer, G. E.~W., and Molenkamp, L.~W., ``Suppression of the persistent spin hall current by defect scattering,'' {\em Phys. Rev. B}~{\bf 70},  041303(R) (2004).

\bibitem{Inoue2006}
Inoue, J.-i., Kato, T., Ishikawa, Y., Itoh, H., Bauer, G. E.~W., and Molenkamp, L.~W., ``Vertex corrections to the anomalous hall effect in spin-polarized two-dimensional electron gases with a rashba spin-orbit interaction,'' {\em Phys. Rev. Lett.}~{\bf 97},  046604 (2006).

\end{thebibliography}
\bibliographystyle{spiebib}

\end{document}